\input harvmac
\def\O{{\cal O}}
\def\nt{\nu_\tau}
\def\nm{\nu_\mu}
\def\ne{\nu_e}

\def\gsim{{~\raise.15em\hbox{$>$}\kern-.85em
          \lower.35em\hbox{$\sim$}~}}
\def\lsim{{~\raise.15em\hbox{$<$}\kern-.85em
          \lower.35em\hbox{$\sim$}~}} 
%%%%%%%%%%%%%%%%%%%%%%%%%%%%
\noblackbox
\baselineskip 14pt plus 2pt minus 2pt
\Title{\vbox{\baselineskip12pt
\hbox{hep-ph/9902293}
\hbox{WIS-99/3/Feb-DPP}}}
{\vbox{
\centerline{Neutrino Masses and Mixing with}
\centerline{Non-Anomalous Abelian Flavor Symmetries}  }}
\centerline{Yosef Nir and Yael Shadmi}
\medskip
\centerline{\it Department of Particle Physics}
\centerline{\it Weizmann Institute of Science, Rehovot 76100, Israel}
\centerline{ftnir@clever.weizmann.ac.il, yshadmi@wicc.weizmann.ac.il}
\bigskip

\baselineskip 18pt
\noindent
The experimental data on atmospheric and solar neutrinos are used to test
the framework of non-anomalous Abelian horizontal gauge symmetries with 
only three light active neutrinos. We assume that the
hierarchy in mass-squared splittings is not accidental and that the small
breaking parameters are not considerably larger than 0.2. We find that the
small angle MSW solution of the solar neutrino problem can only be
accommodated if the $\nm-\nt$ mass hierarchy depends
on the charges of at least three sterile neutrinos. The large angle
MSW solution can be accommodated in simpler models if $\ne$ and
$\nm$ form a pseudo-Dirac neutrino, but it is difficult to induce
large enough deviation from maximal mixing. The vacuum 
oscillation solution can be accommodated rather simply.
We conclude that it is possible to accommodate the neutrino parameters
in the framework of Abelian horizontal symmetries, but it seems that these
parameters by themselves will not provide  convincing evidence for
this framework. 

\Date{2/99}

%%%%%%%%%%%%%%%%%%%%%%%%%%%%
\newsec{Introduction and Results}\
Approximate Abelian horizontal symmetries can explain the smallness
and the hierarchy in the flavor parameters $-$ fermion masses and mixing
angles $-$ in a natural and simple way
\ref\FrNi{C.D. Froggatt and H.B. Nielsen, Nucl. Phys. B147 (1979) 277.}. 
One can think of three types of evidence for such symmetries: First, 
the full theory involves fields that are related to the spontaneous symmetry 
breaking and to the communication of the breaking to the observable sector.
Direct discovery of such particles is, however, very unlikely because 
constraints from flavor changing neutral current (FCNC) processes and from 
Landau poles imply that they should be very heavy
\ref\LNS{M. Leurer, Y. Nir and N. Seiberg,
 Nucl. Phys. B398 (1993) 319, hep-ph/9212278.}.
\nref\DKL{M. Dine, A. Kagan and R. Leigh,
 Phys. Rev. D48 (1993) 4269, hep-ph/9304299.}%
\nref\NiSe{Y. Nir and N. Seiberg,
 Phys. Lett. B309 (1993) 337, hep-ph/9304307.}% 
Second, the supersymmetric flavor parameters are also  determined by the 
selection rules of the horizontal symmetry \refs{\DKL,\NiSe}. (This is likely 
to be the case if supersymmetry  breaking is mediated to the observable sector 
by Planck-scale interactions
\ref\GNR{Y. Grossman, Y. Nir and R. Rattazzi, in {\it Heavy Flavours II},
 eds. A.J. Buras and M. Lindner, Advanced Series on Directions in 
 High Energy Physics, (World Scientific, Singapore), hep-ph/9701231.}; 
in contrast, gauge mediation would erase the effects of the horizontal 
symmetry from the sfermion flavor parameters.) The spectrum of supersymmetric 
particles and, in particular,  supersymmetric effects on FCNC and on 
CP violation could then provide evidence for the horizontal symmetry.  Third, 
it could be that the Yukawa parameters themselves obey simple order of 
magnitude relations that follow from the horizontal symmetry
\ref\LNSb{M. Leurer, Y. Nir and N. Seiberg,
 Nucl. Phys. B420 (1994) 468, hep-ph/9310320.}. In this context, 
Abelian horizontal symmetries have much more predictive power in the lepton 
sector than in the quark sector
\ref\GrNi{Y. Grossman and Y. Nir,
 Nucl. Phys. B448 (1995) 30, hep-ph/9502418.}. 
Neutrino parameters provide then an important input for testing and refining 
this framework.

As concerns neutrino parameters, recent measurements of the flux of 
atmospheric neutrinos (AN) suggest the following mass-squared difference
and mixing between $\nu_\mu$ and $\nu_\tau$
\ref\SKATM{Y. Fukuda {\it et al.}, the Super-Kamiokande Collaboration,
 Phys. Rev. Lett. 81 (1998) 1562, hep-ex/9807003.}:
\eqn\ATM{\Delta m_{23}^2\sim2\times10^{-3}\ eV^2\ ,
\ \ \ \sin^22\theta_{23}\sim1\ .}
On the other hand, measuremnts of the solar neutrino (SN) flux can be explained
by one of the following three options for the parameters of $\nu_e-\nu_x$ 
($x=\mu$ or $\tau$)  oscillations (for a recent analysis, see 
\ref\BKS{J.N. Bahcall, P.I. Krastev and A. Yu. Smirnov,
 Phys. Rev. D58 (1998) 096016, hep-ph/9807216.}):
\eqn\SOL{\matrix{&\Delta m_{1x}^2\ [eV^2]&\sin^22\theta_{1x}\cr
{\rm MSW(SMA)}&5\times10^{-6}&6\times10^{-3}\cr
{\rm MSW(LMA)}&2\times10^{-5}&0.8\cr
{\rm VO}&8\times10^{-11}&0.8\cr}}
Here MSW refers to matter-enhanced oscillations, VO refers 
to vacuum oscillations, and SMA (LMA) stand for small (large) mixing angle.
Only central values are quoted for the various parameters.

Our basic assumption will be that eqs. \ATM\ and \SOL\ imply that the
ratio between the mass splittings is suppressed by the small breaking
parameter of an Abelian horizontal symmetry, $\lambda\sim0.2$, 
while the $\nm-\nt$ mixing angle is not:
\eqn\basic{\eqalign{
\sin\theta_{23}\sim&\ 1,\cr
{\Delta m^2_{12}\over\Delta m^2_{23}}\sim&\ \cases{
\lambda^2-\lambda^4&MSW,\cr \lambda^{10}-\lambda^{12}&VO.\cr}}}
We note, however, that it is not impossible that, if the solar neutrino problem
is a result of the MSW mechanism, the ratio between $\Delta m^2_{\rm SN}$
and $\Delta m^2_{\rm AN}$ is accidentally, rather than parametrically,
suppressed
\nref\BLR{P. Binetruy,  S. Lavignac and P. Ramond,
 Nucl. Phys. B477 (1996) 353, hep-ph/9802334.}%
\nref\ILR{N. Irges,  S. Lavignac and P. Ramond,
 Phys. Rev. D58 (1998) 035003, hep-ph/9802334.}%
\nref\EIR{J.K. Elwood, N. Irges and P. Ramond,
 Phys. Rev. Lett. 81 (1998) 5064, hep-ph/9807228.}%
\nref\ELLN{J. Ellis, G.K. Leontaris, S. Lola and D.V. Nanopoulos,
 hep-ph/9808251.}%
\refs{\BLR-\ELLN}. Then, the analysis of this work and of ref. 
\ref\yyy{Y. Grossman, Y. Nir and Y. Shadmi,
 JHEP 10 (1998) 007, hep-ph/9808355.}\
is irrelevant, and the framework of Abelian horizontal symmetries
can accommodate the neutrino parameters in a simple way.

In a previous work \yyy\ we investigated the implications of \basic\ for 
models where an Abelian horizontal symmetry $H$ is broken by a 
{\it single} small parameter. (For recent related work, see
\nref\BLPR{P. Binetruy, S. Lavignac, S. Petcov and P. Ramond,
 Nucl. Phys. B496 (1997) 3, hep-ph/9610481.}%
\nref\BILR{P. Binetruy, N. Irges, S. Lavignac and P. Ramond,
 Phys. Lett. B403 (1997) 38, hep-ph/9612442.}%
\nref\BHSS{R. Barbieri, L.J. Hall, D. Smith, A. Strumia and N. Weiner,
 JHEP 12 (1998) 017, hep-ph/9807235.}% 
\nref\AlFe{G. Altarelli and F. Feruglio, hep-ph/9809596; hep-ph/9812475.}%
\nref\Jez{M. Jezabek and Y. Sumino,
 Phys. Lett. B440 (1998) 327, hep-ph/9807310.}%
\nref\Eyal{G. Eyal, Phys. Lett. B441 (1998) 191, hep-ph/9807308.}%
\nref\eMa{E. Ma, Phys. Lett. B442 (1998) 238, hep-ph/9807386.}%
\nref\MoNu{R.N. Mohapatra and S. Nussinov,
 Phys. Lett. B441 (1998) 299, hep-ph/9808301.}%
\nref\BHS{R. Barbieri, L.J. Hall and A. Strumia, hep-ph/9808333.}%
\nref\MRS{E. Ma, D.P. Roy and U. Sarkar, Phys. Lett. B444 (1998) 391,
 hep-ph/9810309.}%
\nref\Viss{F. Vissani, JHEP 11 (1998) 025, hep-ph/9810435.}%
\nref\FGN{C.D. Froggatt, M. Gibson and H.B. Nielsen, hep-ph/9811265.}% 
\nref\MaRo{E. Ma and D.P. Roy, hep-ph/9811266.}%
\nref\CCH{K. Choi, E.J. Chun and K. Hwang, hep-ph/9811363.}%
\nref\BeRo{Z. Berezhiani and A. Rossi, hep-ph/9811447.}%
\nref\ShTa{Q. Shafi and Z. Tavartkiladze, hep-ph/9811463.}%
\nref\LiSo{C. Liu and J. Song, hep-ph/9812381.}%
\nref\Tani{M. Tanimoto, hep-ph/9807517; hep-ph/9901210.}%
\refs{\BLPR-\Tani}.) This assumption is best-motivated in models where the 
horizontal symmetry is an anomalous $U(1)_H$ gauge symmetry
\nref\Iban{L. Ibanez, Phys. Lett. B303 (1993) 55.}%
\nref\BiRa{P. Binetruy and P. Ramond,
 Phys. Lett. B350 (1995) 49, hep-ph/9412385.}%
\nref\JaSc{V. Jain and R. Shrock, Phys. Lett. B352 (1995) 83, hep-ph/9412367.}%
\nref\DPS{E. Dudas, S. Pokorski and C.A. Savoy,
 Phys. Lett. B356 (1995) 45, hep-ph/9504292.}%
\nref\YNir{Y. Nir, Phys. Lett. B354 (1995) 107, hep-ph/9504312.}%
\refs{\BiRa-\YNir}. The anomaly is cancelled by the Green-Schwarz mechanism
\ref\GrSc{M. Green and J. Schwarz, Phys. Lett. B149 (1984) 117.}. 
The contribution of the Fayet-Iliopoulos term to the D-term cancels against the contribution from a VEV of a Standard Model (SM) singlet field $S$ with  
$H$-charge that is opposite in sign to $\tr H$. (Without loss of generality, 
we choose the $H$-charge of $S$ to be $-1$). The information about the breaking is communicated to the observable sector (MSSM) at the string scale.
\nref\DSW{M. Dine, N. Seiberg and E. Witten, Nucl. Phys. B289 (1987) 589.}%
\nref\ADS{J. Atick, L. Dixon and A. Sen, Nucl. Phys. B292 (1987) 109.}%
\nref\DIS{M. Dine, I. Ichinoise and N. Seiberg, Nucl. Phys. B293 (1987) 253.}% 
The ratio $\lambda=\vev{S}/m_{\rm Pl}\sim{\tr H\over192\pi^2}$ \refs{\DSW-\DIS} 
provides the small breaking parameter of $H$. The single VEV assumption 
is also plausible if the horizontal symmetry is discrete. In the single VEV 
framework, as explained in \yyy, it is non-trivial to get  large mixing 
together with a large hierarchy as implied by \basic. To obtain that 
one needs to invoke either holomorphic zeros or discrete symmetries, often with
a symmetry group that is a direct product of two factors.

The situation is different if the Abelian horizontal symmetry is
a non-anomalous gauge symmetry. If supersymmetry is not to be broken at the 
scale of spontaneous $H$-breaking, then $H$ should be broken along a D-flat
direction. The simplest possibility then is that two scalars, $S$ and $\bar S$,
of opposite  $H$-charges (say, $\pm1$) assume equal VEVs, $\vev{S}=\vev{\bar S}$
\nref\DLLRS{H. Dreiner, G.K. Leontaris, S. Lola, G.G. Ross and C. Scheich,
 Nucl. Phys. B436 (1995) 461, hep-ph/9409369.}%
\nref\LLR{G.K. Leontaris, S. Lola and G.G. Ross,
 Nucl. Phys. B454 (1995) 25, hep-ph/9505402.}%
\nref\AlNa{C.H. Albright and S. Nandi,
 Phys. Rev. D53 (1996) 2699, hep-ph/9507376.}%
\nref\LLSV{G.K. Leontaris, S. Lola, C. Scheich and J. Vergados,
 Phys. Rev. D53 (1996) 6381, hep-ph/9509351.}%
\nref\Papa{E. Papagregoriu, Zeit. Phys. C64 (1995) 509.}%
\nref\Alla{B.C. Allanach, hep-ph/9806294.}%
\nref\GLLV{M.E. Gomez, G.K. Leontaris, S. Lola and J.D. Vergados,
 hep-ph/9810291.}%
\refs{\DLLRS-\GLLV,\ELLN}. In this work we investigate the implications 
of \basic\ in the two-VEVs framework.

It is straightforward to see that our previous mechanisms are irrelevant in the
new framework. First, with discrete symmetries there is no motivation
for the two-VEVs scenario. There is also no sense in talking about
negative charges. Second, with VEVs of opposite charges there can be
no holomorphic zeros.  On the other hand, this framework is in some sense
less predictive and, consequently, allows new mechanisms to accommodate 
simultaneous large mixing and mass hierarchy between neutrinos. In particular, 
this situation can be obtained naturally with a single
$U(1)$ horizontal symmetry.

As we shall see, if the small-angle MSW solution is found to be valid,
it would mean that the light neutrino masses depend crucially on the
horizontal charges of (at least three) heavy sterile neutrinos.

If, on the other hand, the large-angle MSW solution is confirmed,
two possibilities are allowed.
One is that the light neutrino masses are again affected
by the charges of heavy sterile neutrinos.
The second is that the MSW solution corresponds to oscillations
between the first and second generation neutrinos, which
form a pseudo-Dirac neutrino.
In the latter case, there is another constraint on the neutrino parameters,
$\sin2\theta_{12}<0.9$. This constraint is not simple to satisfy since  
the mixing in the pseudo-Dirac case is close to maximal \BHSS. In fact, 
the deviation from maximal mixing is at most $\O(\lambda^2)$. These models 
are therefore marginally viable. They are only consistent if the 
$\O(\lambda^2)$ correction is enhanced by about three.

The plan of this paper is as follows. In section 2 we specify 
our theoretical framework. In Section 3 we argue that,
in the framework of Abelian horizontal symmetries, $\nm$ and $\nt$
cannot pair to a pseudo-Dirac neutrino. In section 4 we review previous results
in the framework of an anomalous $U(1)$ symmetry. In section 5 we analyze
in detail the lepton parameters in the framework of non-anomalous
Abelian gauge symmetry. (Proofs of some of the statements of this section
are given in the appendix.) We conclude in section 6, emphasizing the
difficulties that the data on solar and atmospheric neutrinos pose
to the framework of Abelian horizontal gauge symmetries.

%%%%%%%%%%%%%%%%%%%%%%%%%%%%
\newsec{The Theoretical Framework}
Our theoretical framework is defined as follows. We consider a low energy
effective theory with particle content that is the same as in the
Supersymmetric Standard Model. In addition to supersymmetry and to the
Standard Model gauge symmetry, there is an approximate $U(1)_H$ symmetry
that is broken by two small parameters $\lambda$ and $\bar\lambda$
\nref\IbRo{L.E. Ibanez and G.G. Ross, Phys. Lett. B332 (1994) 100.}
\refs{\IbRo,\DLLRS-\GLLV,\ELLN}.
The two parameters are assumed to be equal in magnitude:
\eqn\breakH{\lambda=\bar\lambda\sim0.2.}
(The choice of numerical value comes from the quark sector, where the largest
small parameter is $\sin\theta_C=0.22$.) To derive selection rules,
we attribute to the breaking parameters $U(1)_H$ charges:
\eqn\lamcha{H(\lambda)=+1,\ \ \ H(\bar\lambda)=-1.}
Then, the following selection rule applies: {\it Terms in the superpotential
or in the Kahler potential that carry (integer) $H$-charge $n$ are
suppressed by $\lambda^{|n|}$.}\foot{The same selection rule would apply 
in a theory with a single breaking parameter and no supersymmetry.} 

We assume that the active neutrinos (coming from lepton 
doublet supermultiplets $L_i$), are light because of a seesaw mechanism 
involving heavy sterile neutrinos (coming from singlet supermultiplets $N_i$).
However, the relations \basic\ do not necessarily depend
on the charges of all active and sterile neutrinos.
For example, if $\nu_e$ is much lighter than all other neutrinos,
it does not enter \basic. Likewise, if all sterile neutrinos may
be integrated out at the $H$-breaking scale, they do not affect
the effective light neutrino mass matrix.
We will refer to a model containing $n_a$ relevant active neutrinos
and $n_s$ relevant sterile neutrinos as an  $(n_a,n_s)$ model.

We will only consider models where all fields carry integer $H$-charges
(in units of the charge of the breaking parameters). It is possible
that some or all of the lepton fields carry half-integer charges \BHSS.
Then there is a residual, unbroken discrete symmetry.
Such models can be phenomenologically viable and lead to interesting
predictions. We leave the investigation of this class of models to
a future publication
\ref\YaYo{Y. Nir and Y. Shadmi, work in progress.}.

Note that if the horizontal symmetry is a continuous symmetry with
a single breaking parameter, as would be the case for an anomalous $U(1)$,
the selection rule stated above is modified. Superpotential terms
that carry negative $H$-charge cannot appear, as they would require
powers of $\lambda^\dagger$, which is forbidden by holomorphy \LNS.
We refer to these absent terms as ``holomorphic zeros''.

The theory is limited in the sense that it cannot predict the exact 
coefficients of $\O(1)$ for the various terms. Wherever we use the symbol 
``$\sim$" below we mean to say that the unknown coefficients of $\O(1)$ are 
omitted.

RGE effects could enhance the neutrino mixing angle
\nref\ChPl{P.H. Chankowski and Z. Pluciennik,
 Phys. Lett. B316 (1993) 312, hep-ph/9306333.}%
\nref\BLP{K.S. Babu, C.N. Leung and J. Pantaleone,
 Phys. Lett. B319 (1993) 191, hep-ph/9309223.}%
\nref\TaRG{M. Tanimoto, Phys. Lett. B360 (1995) 41, hep-ph/9508247.}%
\nref\HaMa{N. Haba and T. Matsuoka,
 Prog. Theor. Phys. 99 (1998) 831, hep-ph/9710418.}% 
\nref\HOS{N. Haba, N. Okamura and M. Sugiura, hep-ph/9810471.}%
\refs{\ChPl-\HOS,\ELLN}. The enhancement can take place if $\tan\beta$
is large and if the mass ratio between the corresponding neutrinos is not 
small. These enhancement effects are not important in our framework
and we will not take them into account.

%%%%%%%%%%%%%%%%%%%%%%%%%%%%
\newsec{On Pseudo-Dirac Neutrinos}
As a first step in our discussion,
we would like to make a general comment about pseudo-Dirac neutrinos
in the framework of Abelian horizontal symmetries. In many of our examples 
(here and in \yyy), two of the three active neutrinos pair to
form a pseudo-Dirac neutrino. In all of these examples, the parameters
of the pseudo-Dirac neutrino (maximal mixing and very small mass splitting)
are fitted to solve the SN problem. Since AN observations seem to favor
{\it maximal}, and not just generic ${\cal O}(1)$, mixing, one may 
wonder whether we could find a model where, indeed, the mass splitting between 
the components of the pseudo-Dirac neutrino corresponds to 
$\Delta m^2_{\rm AN}$. We argue now that this is impossible.

The argument goes as follows. Let us define $m_{\rm pD}$ to be the mass of 
the pseudo-Dirac neutrino and $\delta_{\rm pD}\ll m_{\rm pD}$ to be the mass 
splitting between its components. An Abelian symmetry cannot give an exact 
relation between three entries in the mass matrix. (The symmetric structure 
of  $M_{\nu}$ relates pairs of entries, which enables us to find models with 
a pseudo-Dirac neutrino.) Therefore, the mass-squared splitting between the 
pseudo-Dirac neutrino and the other mass eigenstate is at least 
${\cal O}(m_{\rm pD}^2)$.\foot{It is of course possible to fine tune the 
$\O(1)$ coefficients to get a stronger degeneracy \LLSV.} On the other hand, 
the mass-squared splitting between the components of the pseudo-Dirac 
neutrino is ${\cal O}(\delta_{\rm pD}m_{\rm pD})$. Since $\delta_{\rm pD}\ll 
m_{\rm pD}$, the mass-squared splitting between the components of the 
pseudo-Dirac neutrino is much smaller than the mass-squared splitting 
between the pseudo-Dirac neutrino and the third mass eigenstate. Therefore, 
the former corresponds to $\Delta m^2_{\rm SN}$ and the latter to 
$\Delta m^2_{\rm AN}$.\foot{If one considers {\it only} the AN problem, then 
it is of course possible that $\nm$ and $\nt$ form a pesudo-Dirac neutrino 
\Alla.}

It is worth emphasizing that the discussion of this section applies
to any Abelian symmetry, continuous or discrete.

%%%%%%%%%%%%%%%%%%%%%%%%
\newsec{Continuous Symmetry with a Single Breaking Parameter}
Before moving on to our discussion of continuous symmetries with
two breaking parameters, let us recall some results of \yyy\
in the single VEV framework. 

The main obstacle in obtaining, within the single-VEV framework, large mixing
between hierarchically separated neutrinos, can be explained as follows.
Consider a single $U(1)_H$ symmetry. Large mixing between, say, $\nu_2$
and $\nu_3$ can only be obtained in two cases: either the $H$-charges of
the lepton doublets are equal \GrNi, $H(L_2)=H(L_3)$, or they are opposite
\BLPR, $H(L_2)=-H(L_3)$. In the first case the mixing is $\O(1)$ but the masses
are of the same order of magnitude, $m(\nu_2)\sim m(\nu_3)$. In the second
case, to a good approximation, the mixing is maximal, $\sin^22\theta_{23}=1$ 
and the masses are equal, $m(\nu_2)=m(\nu_3)$. (This is the case of a 
pseudo-Dirac neutrino.) In either case, there is no mass hierarchy.
If the symmetry is continuous but more complicated, say $U(1)_1\times U(1)_2$
with the respective breaking parameters of order $\lambda^m$ and $\lambda^n$,
then one can still define an {\it effective $H$-charge},
\eqn\effH{H_{\rm eff}=mH_1+nH_2.}
Large mixing can only be obtained for $H_{\rm eff}(L_2)=\pm H_{\rm eff}(L_3)$,
so that, again, there is no mass hierarchy.
This conclusion can only be evaded if holomorphic zeros appear
in the neutrino mass matrix such that one of the two mass eigenvalues vanishes.
In other words, in the single VEV framework, when considering
$\nu_2$ and $\nu_3$ only, the only way to get large
mixing and a large hierarchy between them is to make $\nu_2$
massless. Clearly though, we need to generate a mass for $\nu_2$
as well in order to account for \basic. This can be arranged by having 
$\nu_2$ combine with $\nu_1$ to form a pseudo-Dirac neutrino.
But then $\sin 2\theta_{12}$ is large, and we cannot obtain the
small angle MSW solution.

In the two VEV framework, the notion of effective charge is not very useful.
It is the absence of similarly powerful constraints that, on the one hand,
makes the two VEV models less predictive but, on the other hand, allows
one to accommodate the neutrino parameters more easily.

%%%%%%%%%%%%%%%%
%%%%%%%%%%%%%%%%%%%%%%%%
\newsec{Continuous Symmetry with Two Breaking Parameters}
%%%%%%%%%%%%%%%%
\subsec{No pseudo-Dirac neutrino}
We start our investigation of Abelian horizontal symmetries with two
breaking parameters by assuming that the solution of the SN problem 
does not involve a pseudo-Dirac neutrino. Then the hierarchy
of mass-squared splittings is simply related to the hierarchy of masses:
\eqn\masqma{{\Delta m^2_{12}\over\Delta m^2_{23}}\sim{m_2^2\over m_3^2}\ .}
The case $m_2\ll m_1$ is only relevant to the VO solution of the SN problem 
and does not affect our conclusions in this subsection.  
These assumptions allow us to 
investigate the relevant parameters in a $(2,n_s)$ framework.

A simple mechanism for inducing large mixing between hierarchically 
separated masses can be demonstrated in a simple $(2,0)$ model. In 
this model, the lepton mass matrices have the form:
\eqn\Mnu{M_\nu\sim{\vev{\phi_u}^2\over M}\pmatrix{
\lambda^{2|H(L_2)|}&\lambda^{|H(L_2)+H(L_3)|}\cr
\lambda^{|H(L_2)+H(L_3)|}&\lambda^{2|H(L_3)|}\cr},}
\eqn\Mel{M_{\ell^\pm}\sim\vev{\phi_d}\pmatrix{
\lambda^{|H(L_2)+H(\bar\ell_2)|}&\lambda^{|H(L_2)+H(\bar\ell_3)|}\cr
\lambda^{|H(L_3)+H(\bar\ell_2)|}&\lambda^{|H(L_3)+H(\bar\ell_3)|}\cr}.}
To have $m_2\ll m_3$, we need $|H(L_2)|\neq|H(L_3)|$. We can still get
a mixing of $\O(1)$ if
\eqn\mixell{H(L_2)+H(L_3)=-2H(\bar\ell_3).}
In a general $(n_a,n_s)$ model, large mixing could arise also from the
neutrino Dirac mass matrix or the Majorana mass matrix of the sterile
neutrinos. But in all these cases, it is required, similarly to
\mixell, that
\eqn\evendiff{H(L_2)-H(L_3)=0({\rm mod}\ 2).}

Eq. \evendiff\ has interesting implications 
for the mass hierarchy. From eq. \Mnu\ we learn that
\eqn\toolarge{{m_2^2\over m_3^2}\sim\lambda^{4[|H(L_2)+H(L_3)|-2|H(L_3)|]}
\sim\lambda^{8n}\ \ \ (n=\ {\rm integer}).}
This creates a phenomenological
problem. If the hierarchy is $\lambda^0$, $\Delta m^2_{\rm SN}$ is too large.
The next weakest possibilities are $\lambda^8$ or $\lambda^{16}$. But for the 
MSW solutions we need $\lambda^{2-4}$ and for the VO solution we need  
$\lambda^{10-12}$. We can achieve neither in this framework. The MSW solutions
are particularly disfavored; the VO solution may still correspond to the 
$\lambda^8$ hierarchy if $\lambda$ is actually close to 0.1 (rather than the 
value of 0.2 that we usually use).\foot{The MSW solution can still be
accommodated if the small breaking parameters are large, $\lambda\sim0.5$ 
\LLR.}

Of course, we have only considered the (2,0) framework. However, one can
show that eq. \toolarge\ holds also in the (2,2) case. Within ($2,n_s\geq3$)
models, however, the hierarchy is an integer power of $\lambda^4$
and, consequently, could be milder.~\foot{We thank F. Feruglio
for pointing out a $(2,3)$ example with $\lambda^4$ hierarchy.} 
A proof of these statements can be
found in the Appendix. We conclude then
that one of the following options has to happen in order to achieve neutrino 
parameters that are consistent with the MSW solutions:
\item{(i)} There are at least three sterile neutrinos that affect the
mass hierarchy $m_2/m_3$. (A $(2,3)$ model which accommodates the MSW(SMA) 
solution can be found in \AlFe.)
\item{(ii)} $\ne$ and $\nm$ form a pesudo-Dirac neutrino. This case is,
of course, relevant only to the MSW(LMA) solution.

A simple example of how the VO solution can be implemented in this
framework,
%!
with $\lambda\sim 0.1$, 
 is achieved with the following set of charges:
\eqn\LVOone{\eqalign{
&L_1(+9),\ \ \ L_2(+3),\ \ \ L_3(+1),\cr
&\ell_1(-15),\ \ \ \ell_2(-6),\ \ \ \bar\ell_3(-2).\cr}}
The neutrino mass matrix is
\eqn\MnuVOone{M_\nu\sim{\vev{\phi_u}^2\over M}\pmatrix{
\lambda^{18}&\lambda^{12}&\lambda^{10}\cr\lambda^{12}&\lambda^6&\lambda^4\cr
\lambda^{10}&\lambda^4&\lambda^2\cr},}
which yields
\eqn\SAVOone{{\Delta m^2_{\rm SN}\over\Delta m^2_{\rm AN}}\sim\lambda^{8}.}
For the charged lepton mass matrix we find 
\eqn\MellVOone{M_\ell\sim\vev{\phi_d}\pmatrix{
\lambda^6&\lambda^3&\lambda^7\cr\lambda^{12}&\lambda^3&\lambda\cr
\lambda^{14}&\lambda^5&\lambda\cr},}
which gives $\sin\theta_{23}\sim1$ and, for $\tan\beta\sim\lambda^{-2}$,
the required charged lepton mass hierarchy. 

%%%%%%%%%%%%%%%%
\subsec{A pseudo-Dirac neutrino: Hierarchy of mass splittings without hierarchy of masses}
A large $\nm-\nt$ mixing, relevant to the AN problem, and a large $\ne-\nm$ 
mixing, relevant to the SN problem, could arise from very different
mechanisms: $\O(1)$ $\nm-\nt$ mixing from unequal charges (as discussed
in the previous subsection), and maximal $\ne-\nm$ mixing from their 
pairing to a pseudo-Dirac combination. Such a situation opens up the 
interesting possibility that there is actually no mass hierarchy: All three 
neutrino masses may be of the same order of magnitude, which is the scale set 
by AN, with the mass-squared {\it splittings} hierarchically separated.

It is simple to see that all three neutrino masses are of the same order of 
magnitude if we take  
\eqn\nohi{|H(L_1)+H(L_2)|=2|H(L_3)|.} 
Instead of \masqma\ we now have  
\eqn\masqma{{\Delta m^2_{12}\over\Delta m^2_{23}}\sim
{m_{\rm pD}\delta_{\rm pD}\over m_3^2}\sim{\delta_{\rm pD}\over m_{\rm pD}}.}
The dependence of this ratio on the lepton charges is given by
\eqn\nohipD{{\Delta m^2_{\rm SN}\over\Delta m^2_{\rm AN}}\sim
\lambda^{2[|H(L_2)|-|H(L_3)|]}\lsim\lambda^4.} 
In contrast to \toolarge, the mass hierarchy is appropriate for
the MSW parameters. Yet, the MSW(LMA) solution cannot
be achieved because the deviation from maximal mixing is suppressed by
at least $\O(\lambda^4)$. The reason is that a deviation of $\O(\lambda^2)$ can be achieved only if $H(L_1)-H(L_2)$ is odd, but this is impossible because
of \nohi. Therefore, in this class of models, only the VO 
solution can be accommodated. We now demonstrate this by an explicit example.
 
Consider the following set of $H$-charges for the lepton fields in the (3,0)
framework:
\eqn\Lbcharge{\eqalign{
L_1(+7),\ \ \ L_2(-5),&\ \ \ L_3(+1),\cr
\bar\ell_1(-15),\ \ \ \bar\ell_2(+10),&\ \ \ \bar\ell_3(+2).\cr}}
The neutrino mass matrix is of the form
\eqn\MnunoH{M_\nu\sim{\vev{\phi_u}^2\over M}
\pmatrix{\lambda^{14}&A\lambda^2&\lambda^8\cr
A\lambda^2&\lambda^{10}&\lambda^4\cr \lambda^8&\lambda^4&B\lambda^2\cr}.}
For later purposes we explicitly wrote down the ${\cal O}(1)$ coefficients,
$A$ and $B$, of the dominant entries. We see that all three neutrinos have
masses of the same order of magnitude, that is
\eqn\numass{m(\nu_i)\sim{\vev{\phi_u}^2\over M}\lambda^2\ \ \ {\rm for}\ 
i=1,2,3.}
The mass splittings are, however, hierarchical:
\eqn\splitH{{\Delta m^2_{12}\over\Delta m^2_{23}}\sim\lambda^8,}
which fits the VO solution for $\lambda\sim0.1$. 
A large $2-3$ mixing is obtained from the charged lepton sector:
\eqn\MlenoH{M_\ell\sim\vev{\phi_d}\pmatrix{\lambda^8&\lambda^{17}&\lambda^9\cr
\lambda^{20}&\lambda^{5}&\lambda^3\cr \lambda^{14}&\lambda^{11}&\lambda^3\cr}.}
We learn that 
\eqn\mixnoH{\sin^22\theta_{12}\simeq1,\ \ \ \sin\theta_{13}\sim\lambda^6,\ \ \ 
\sin\theta_{23}\sim1.}

Note that this scenario of hierarchical mass-squared splittings is not 
included in ref. \BHSS. The form of the neutrino mass matrix \MnunoH\
in the charged lepton mass basis is given to a good approximation by
\eqn\Mnuellm{M_\nu\sim{\vev{\phi_u}^2\over M}\lambda^2\pmatrix{0&Ac&As\cr
Ac&Bs^2&Bcs\cr As&Bcs&Bc^2\cr},}
where $c\equiv\cos\theta_{23}$ and $s\equiv\sin\theta_{23}$. Indeed, this 
corresponds to neither of the two forms advocated in \BHSS. It is amusing 
to note, however, that it can be presented as the sum of these two forms.

%%%%%%%%%%%%%%%%%%%%%%%%%%%%
\subsec{Large mixing from equal effective charges}
A different way of obtaining large mixing together with a
large hierarchy is the analog of the `holomorphic zeros' mechanism
of ref. \yyy. In both frameworks we take the horizontal symmetry
to be $U(1)_1\times U(1)_2$, with equal effective charges (see eq. \effH)
for $L_2$ and $L_3$, so that the $2-3$ mixing is ${\cal O}(1)$.
The separate charges can, however, be chosen so as to induce a holomorphic
zero in the single VEV framework and to suppress one of the masses in the
two VEV framework.
 
As an example consistent with the MSW(LMA) solution,
consider the following set of charges for the lepton fields within
a (3,0) model \yyy:
\eqn\Lccharge{\eqalign{
L_1(1,0),\ \ \ L_2(-1,1),&\ \ \ L_3(0,0),\cr
\bar\ell_1(3,4),\ \ \ \bar\ell_2(3,2),&\ \ \ \bar\ell_3(3,0).\cr}}
The lepton mass matrices are of the form
\eqn\MnuHol{M_\nu\sim{\vev{\phi_u}^2\over M}
\pmatrix{\lambda^{2}&\lambda&\lambda\cr
\lambda&\lambda^4&\lambda^2\cr \lambda&\lambda^2&1\cr},\ \ \ 
M_{\ell^\pm}\sim\vev{\phi_d}\pmatrix{\lambda^8&\lambda^6&\lambda^4\cr
\lambda^7&\lambda^5&\lambda^3\cr\lambda^7&\lambda^5&\lambda^3\cr}.}
Without the positively charged $\bar\lambda_1$, the (22), (23) and (32)
entries in $M_\nu$
would vanish because of holomorphy \yyy. Here, the holomorphic zeros
are lifted, but the new entries are small and affect neither the mass
hierarchy nor the mixing. Thus, the analysis of \yyy\ is still valid, yielding
\eqn\eqeff{{\Delta m^2_{12}\over\Delta m^2_{23}}\sim\lambda^3,\ \ 
\sin2\theta_{12}=1-\O(\lambda^2),\ \ s_{23}\sim1,\ \ s_{13}\sim\lambda.}

The VO solution is similarly obtained with
\eqn\Lcchargevo{\eqalign{
L_1(1,-4),\ \ \ L_2(-2,2),&\ \ \ L_3(0,0),\cr
\bar\ell_1(6,5),\ \ \ \bar\ell_2(3,2),&\ \ \ \bar\ell_3(3,0).\cr}}
This gives
\eqn\MnuHolvo{M_\nu=\pmatrix{\lambda^{10}&\lambda^3&\lambda^5\cr
\lambda^3&\lambda^8&\lambda^4\cr \lambda^5&\lambda^4&1\cr},}
yielding ${\Delta m^2_{12}\over\Delta m^2_{23}}\sim\lambda^{11}$.

%%%%%%%%%%%%%%%%%%
\newsec{Conclusions}
In models with a non-anomalous Abelian horizontal gauge symmetry, one
expects that the symmetry is spontaneously broken by fields of opposite
horizontal charges that acquire equal VEVs. We find two general mechanisms
by which such models can accommodate the neutrino parameters that explain both
the atmospheric neutrino and the solar neutrino problems. In particular,
these mechanisms allow for
\eqn\general{\sin\theta_{23}\sim1,\ \ \ \ 
{\Delta m^2_{12}\over \Delta m^2_{23}}\ll1.}

The possible scenarios divide into two general classes:
\item{(i)} The three neutrino masses are hierarchical. Then the hierarchy
between the mass-squared splittings of eq. \general\ is an integer power
of $\lambda^4$. The bound can only be saturated in models where at
least three sterile neutrinos affect the mass hierarchy. Otherwise,
the hierarchy is an integer power of $\lambda^8$ which is inconsistent with
the MSW solutions. The VO solution can be accommodated in the
simple models that have $\lambda^{8n}$ hierarchy
if the small breaking parameter is somewhat smaller than
the `canonical' value of $0.2$ related to the Cabibbo mixing.
\item{(ii)} The two lighter neutrinos form a pseudo-Dirac neutrino.
The mixing related to the solar neutrino solution is then close to maximal,
so that obviously only large angle solutions to the SN problem are possible.
It is not simple, though not impossible, to obtain large enough deviation
from maximal mixing as necessary for the large angle MSW solution.
The VO solution is, again, simply accommodated.
A special case in this class is that of three same order-of-magnitude
neutrino masses, with hierarchical splittings.

We emphasize that our arguments are not valid for {\it discrete}
Abelian symmetries \yyy. Moreover, they can be circumvented even if the
symmetry is continuous but the spontaneous symmetry breaking is not complete,
leaving a residual exact discrete symmetry \refs{\BHSS,\YaYo}. 
In particular, it has been demonstrated that the MSW(SMA) solution can be
easily generated in these cases \refs{\yyy,\BHSS}.

While the conclusion of both this work and  of ref.~\yyy\ is that
large mixing and large hierarchy can be accommodated in the framework of
Abelian horizontal symmetries, we would still like to emphasize the 
following points:
\item{a.} The most predictive class of models of Abelian horizontal
symmetries is that of an anomalous $U(1)$ with holomorphic zeros
having no effect on the physical parameters \GrNi. The various solutions
suggested here and in \yyy\ require that either the symmetry is
non-anomalous with two breaking parameters, or the symmetry is discrete,
or that holomorphic zeros do play a role. In all these cases there
is a loss of predictive power. If, indeed, \general\ holds in nature,
it would mean that neutrino parameters by themselves will not make
a convincing case for the Abelian horizontal symmetry idea, even if they
cannot rule it out.
\item{b.} We argued here that, if \general\ holds, the neutrino parameters
that correspond to the atmospheric neutrino oscillations are not related
to a pseudo-Dirac neutrino. Consequently, while Abelian horizontal
symmetries allow for $\O(1)$ $\nm-\nt$ mixing, they cannot explain {\it maximal}
mixing (except as an accidental result). If the case for $\sin^22\theta_{23}
=1$ is experimentally made with high accuracy, and the solar
neutrino problem is indeed solved by neutrino oscillations, the
framework of Abelian horizontal symmetries would become less attractive.

\bigskip
\noindent
{\bf Acknowledgements}
\smallskip
\noindent
We are grateful to Ferruccio Feruglio for providing us with
a counter-example to statements made in the original version
of this paper. We thank Enrico Nardi for helpful discussions.
This work was supported in part by the
United States $-$ Israel Binational Science Foundation (BSF) 
and by the Minerva Foundation (Munich).

%%%%%%%%%%%%
%%%%%%%%%%%%
\appendix{A}{Mass Hierarchy with Large Mixing}
We study models with $n_a$ active neutrinos and $n_s$ sterile ones
(`($n_a,n_s$) models'). We will argue that, as concerns the implications 
for $\Delta m^2_{12}$, the models divide into three classes:
\item{a.} Effective $(2,n_s\geq3)$ models, where we find that
$\Delta m^2_{12}/\Delta m^2_{23}\sim\lambda^{4n}$ ($n=$ integer).
\item{b.} Effective $(2,2)$ models, where we find that
$\Delta m^2_{12}/\Delta m^2_{23}\sim\lambda^{8n}$ ($n=$ integer).
\item{c.} Models where $\ne$ and $\nm$ form a pseudo-Dirac neutrino,
where $\Delta m^2_{12}\ll m^2_{1,2}$.

The first step in our argument is to show that the large mixing between 
$\nu_2$ and $\nu_3$ requires that the horizontal charges of $L_2$ and 
$L_3$ obey
\eqn\evendif{H(L_2)-H(L_3)=0({\rm mod}\ 2).}
There are three possible sources (in the interaction basis) for large
mixing: (i) the charged lepton mass matrix $M_\ell$, (ii) the neutrino 
Dirac mass matrix $M_\nu^{\rm Dir}$, and (iii) the sterile neutrino Majorana
mass matrix $M_{\nu_s}^{\rm Maj}$. We now examine the three mechanisms in turn.

(i) If the large mixing arises from $M_\ell$, then 
\eqn\Dirmix{(M_\ell)_{23}\sim(M_\ell)_{33}\Longrightarrow
|H(L_2)+H(\bar\ell_3)|=|H(L_3)+H(\bar\ell_3)|\ .}
Therefore, either $H(L_2)=H(L_3)$, or $H(L_2)+H(L_3)=-2H(\bar\ell_3)$.
In either case, \evendif\ holds.

(ii) If the large mixing arises from $M_\nu^{\rm Dir}$, then
a similar argument holds with $H(\bar\ell_3)$ replaced by $H(N_3)$.

(iii) Let us examine the conditions for large mixing induced by 
$M_{\nu_s}^{\rm Maj}$. To do so, we consider a (2,2) model.
We define the relevant matrix elements by
\eqn\Mnus{(M_{\nu_s}^{\rm Maj})^{-1}=
\pmatrix{r_{22}&r_{23}\cr r_{23}&r_{33}\cr}.}
For simplicity, we take $M_\nu^{\rm Dir}$ to be diagonal (consistent with
our assumption that the large mixing does not come from this matrix):
\eqn\Mdird{M_\nu^{\rm Dir}=\pmatrix{d_2&\cr &d_3\cr}.}
Then
\eqn\Mlight{M_{\nu_a}^{\rm Maj}=(M_\nu^{\rm Dir})(M_{\nu_s}^{\rm Maj})^{-1}
(M_\nu^{\rm Dir})^T=\pmatrix{d_2^2r_{22}&d_2d_3r_{23}\cr
d_2d_3r_{23}&d_3^2r_{33}\cr}.}
Large mixing can be induced in two cases: first, $d_2d_3r_{23}\gg d_3^2r_{33}$
which leads to a pesudo-Dirac neutrino in contrast to our assumptions;
second, $d_2d_3r_{23}\sim d_3^2r_{33}$, which can be achieved with
\eqn\condMs{{d_2r_{23}\over d_3r_{33}}\sim\lambda^{
|H(L_2)+H(N_2)|-|H(L_3)+H(N_3)|+|H(N_2)+H(N_3)|-2|H(N_2)|}\sim1.}
The condition on the exponent is then of the form
\eqn\expo{a_2H(L_2)+a_3H(L_3)+2b_2H(N_2)+2b_3H(N_3)=0,}
where $a_2,a_3=\pm1$, $b_3=0,\pm1$ and $b_2=0,\pm1,\pm2$.
Clearly, it leads to \evendif.

The second step is to find the hierarchy between the masses in terms
of the lepton charges. If no pair among the active neutrinos forms
a pseudo-Dirac state, then we can estimate the mass ratio from
\eqn\masrat{{m_2\over m_3}\sim{\det M^{(n_s+2)}\det M^{(n_s)}\over
[\det M^{(n_s+1)}]^2}.}
Here, $\det M^{(n_s)}$ is the product of the masses of the sterile
neutrinos, which is approximately equal to $\det M^{\rm Maj}_{\nu_s}$, and
$\det M^{(n_s+n^\prime_a)}$  is the product of the masses of the sterile
neutrinos and the masses of the $n^\prime_a$ heaviest active neutrinos.
To  estimate the masses, we use 
\eqn\MnuDir{(M_\nu^{\rm Dir})_{ij}\sim\vev{\phi_u}\lambda^{|H(L_i)+H(N_j)|},}
\eqn\MnuMaj{(M_{\nu_s}^{\rm Maj})_{ij}\sim M\lambda^{|H(N_i)+H(N_j)|}.}

In $\det M^{(n_s)}$, each $H(N_i)$ appears twice in the exponent, each
time with either a plus or a minus sign. Consequently, we obtain the following
type of dependence on lepton charges:
\eqn\detns{\det M^{(n_s)}\sim M^{n_s}\lambda^{a_iH(N_i)},\ \ \ a_i=0,\pm2\ .}
Similarly, in $\det M^{(n_s+1)}$, each $H(N_i)$ appears twice in the exponent, 
each time with a plus or a minus sign. As concerns $H(L_i)$, there are
two possibilities: either a single $H(L_i)$ appears twice, each time 
with either a 
plus or a minus sign, or two different charges appear once:
\eqn\detnso{\eqalign{\det M^{(n_s+1)}\sim&\  M^{n_s-1}\vev{\phi_u}^2
{\rm max}\left\{\lambda^{a_iH(N_i)+bH(L_j)},
\lambda^{a_iH(N_i)+c[H(L_j)+dH(L_k)]}\right\}\cr
&\ \ \ a_i,b=0,\pm2,\ c,d=\pm1.\cr}}
In a similar manner, we obtain
\eqn\detnst{\det M^{(n_s+2)}\sim M^{n_s-2}\vev{\phi_u}^4
\lambda^{a_iH(N_i)+b_jH(L_j)},\ \ \ a_i,b_j=0,\pm2.}
It is straightforward to see that each of $\det M^{(n_s)}$, 
$\det M^{(n_s+2)}$ and $[\det M^{(n_s+1)}]^2$ depends on $\lambda^{2n}$
where $n$ is an integer.

We remind the reader that eq. \masrat\ holds only when
there is no pseudo-Dirac light neutrino. In other words, it holds
in effective $(2,n_s)$ models. Then, $H(L_1)$ does not appear 
in eqs. \detns-\detnst; in particular, the $H(L_i)$-dependence of the 
second factor in \detnso\ is of the form $c[H(L_2)+dH(L_3)]$. Eq. \evendif\ 
implies that this combination of charges is even. Generically, however, 
this fact has no special consequences and we find
\eqn\genhie{{\Delta m^2_{12}\over \Delta m^2_{23}}\sim{m_2^2\over m_3^2}
\sim\lambda^{4n}\ \ \ [(2,n_s)\ {\rm models}].}

The situation is more constrained if the charges of only two of the 
sterile neutrinos (say, $N_2$ and $N_3$) affect $s_{23}$ and $m_2/m_3$. 
This is the $(2,2)$ model. There are two additional
special features in this case:
\item{(i)} For $n_s=2$, we have $\det M^{(n_s=2)}\sim M^2
\lambda^{2|H(N_2)+H(N_3)|}$
which, in particular, allows only $a_i=\pm2$ in \detns.
\item{(ii)} For any $(n,n)$ model, we have $\det M_\nu=
[\det M_\nu^{\rm Dir}]^2$. Consequently, $a_i,b_j=\pm2$ (and cannot
vanish) in eq.\detnst.

As a result of all these features, we now find
\eqn\tthie{{\Delta m^2_{12}\over \Delta m^2_{23}}\sim{m_2^2\over m_3^2}
\sim\lambda^{8n}\ \ \ [(2,2)\ {\rm models}].}
 
\listrefs
\end